\documentclass[twoside,leqno,twocolumn]{article}  
\usepackage{ltexpprt} 
\usepackage{balance}
\usepackage[T1]{fontenc}
\usepackage[latin1]{inputenc}
\usepackage{times}
\newcommand{\komment}[1]{}
\usepackage{helvet}
\usepackage{amssymb}
\usepackage{algorithm}
\usepackage[noend]{algorithmic} 
\usepackage{courier}
\usepackage{graphicx,color}  
\newcommand{\openbox}{\leavevmode
  \hbox to.77778em{%
  \hfil\vrule
  \vbox to.675em{\hrule width.6em\vfil\hrule}%
  \vrule\hfil}}

\newcommand{\squishlist}{
   \begin{list}{$\bullet$}
    { \setlength{\itemsep}{0pt}      \setlength{\parsep}{3pt}
      \setlength{\topsep}{3pt}       \setlength{\partopsep}{0pt}
      \setlength{\leftmargin}{1.5em} \setlength{\labelwidth}{1em}
      \setlength{\labelsep}{0.5em} } }
\newcommand{\squishlisttwo}{
   \begin{list}{$\bullet$}
    { \setlength{\itemsep}{0pt}    \setlength{\parsep}{0pt}
      \setlength{\topsep}{0pt}     \setlength{\partopsep}{0pt}
      \setlength{\leftmargin}{2em} \setlength{\labelwidth}{1.5em}
      \setlength{\labelsep}{0.5em} } }
\newcommand{\squishend}{
    \end{list}  }
\def\standardlength{6cm}
\begin{document}

	 \title{\Large A Better Alternative to Piecewise Linear Time Series Segmentation\thanks{Supported by
  	 NSERC grant 261437.}}
%
%

%

%
\author{Daniel Lemire\thanks{Universit\'e du Qu\'ebec \`a Montr\'eal (UQAM)}}
\date{}
\maketitle

%
%
\maketitle
\begin{abstract}
Time series are difficult to monitor, summarize and predict.
Segmentation organizes time series into 
few intervals having uniform characteristics 
(flatness, linearity, modality, monotonicity and so on).
For scalability, we require fast linear time algorithms.
The popular piecewise linear model can determine where the
data goes up or down and at what rate. Unfortunately, when
the data does not follow a linear model, the computation
of the local slope creates  overfitting. 
We propose an adaptive time series model where the polynomial
degree of each interval  vary (constant, linear and so on).  Given a number of
regressors, the cost of each interval is  its polynomial degree: constant intervals
cost  1~regressor, linear intervals cost 2 regressors, and so on. 
Our goal is to minimize the Euclidean ($l_2$) error for a given model complexity.
Experimentally, we investigate the model where intervals can be either constant
or linear.
Over synthetic random walks, 
historical stock market prices,
and electrocardiograms, the adaptive model  provides a more accurate segmentation
than the  piecewise linear model without increasing the cross-validation error or
the running time, while
providing a richer vocabulary to applications.
Implementation issues, such as numerical stability and real-world performance, are discussed.
\end{abstract}



\section{Introduction}


Time series are ubiquitous in finance,
engineering, and science. They are an application area of growing importance in database research~\cite{lowellreport}.
Inexpensive sensors can record data points at 5~kHz or more, generating one million samples
every three minutes.
The primary purpose of time series segmentation is dimensionality reduction.
It is 
 used to find frequent patterns~\cite{han98} or classify time series~\cite{KeoghP98}. 
Segmentation points divide the time axis into intervals behaving approximately according to a simple model.
Recent work on segmentation used quasi-constant or quasi-linear intervals~\cite{657889}, 
quasi-unimodal intervals~\cite{Haiminen04}, step, ramp or impulse~\cite{citeulike:799461}, or quasi-monotonic intervals~\cite{YLBIJCAI05,AAAI05}.  A good time series segmentation algorithm must 
\squishlist
 \item be fast (ideally run in linear time with low overhead);
\item provide the necessary vocabulary such as flat, increasing at rate $x$, decreasing at rate $x$, \ldots;
\item be accurate (good model fit and cross-validation).
\squishend

\begin{figure}[h]
	\center
\includegraphics[width=5.45cm,angle=270]{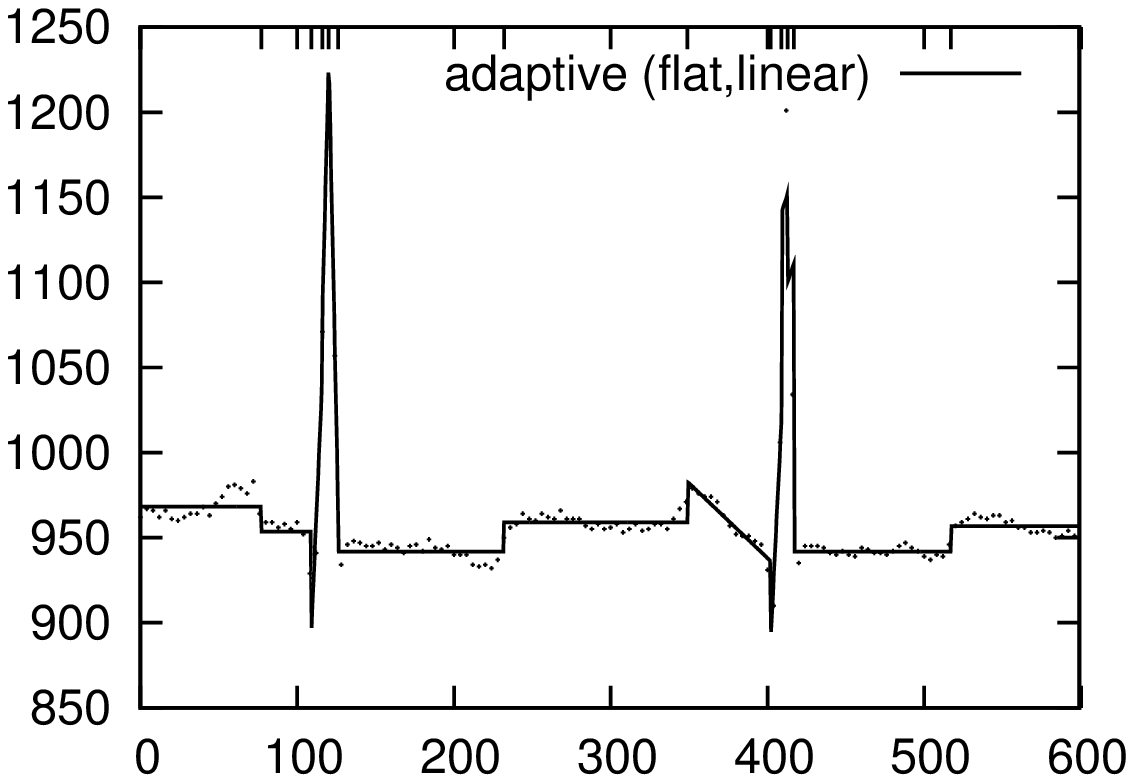}
	\center
\includegraphics[width=5.45cm,angle=270]{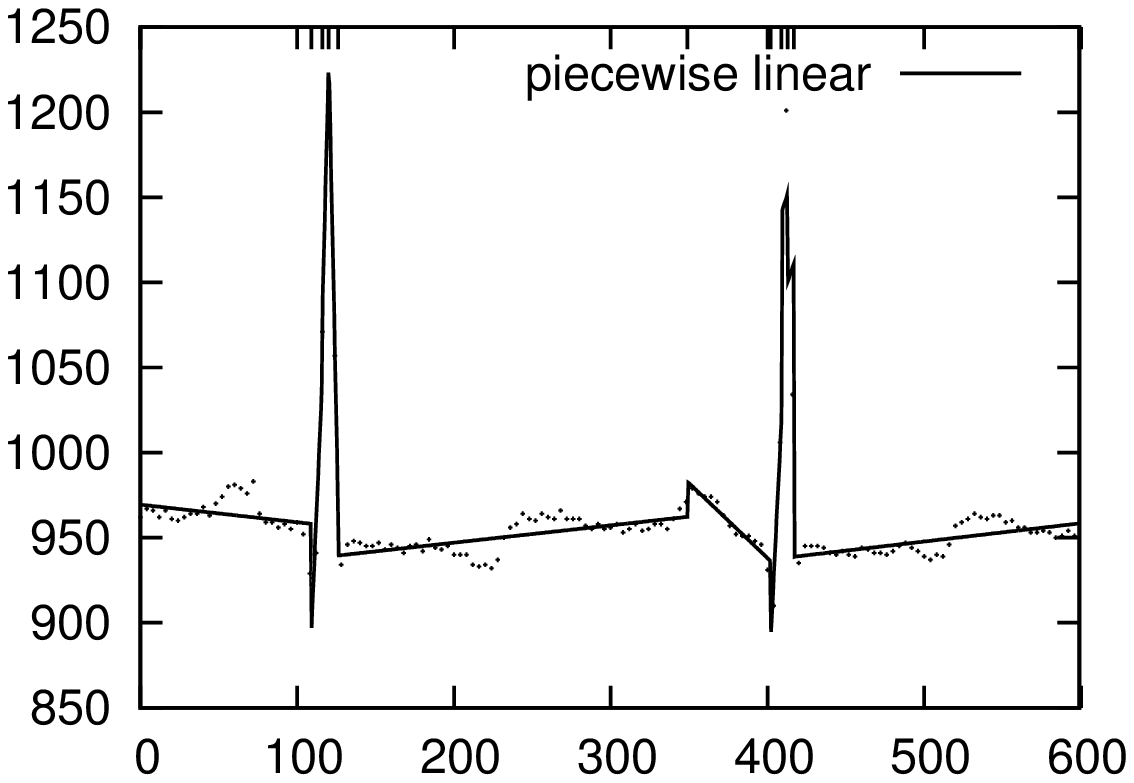}
\caption{\label{sunwplot}Segmented electrocardiograms (ECG) with adaptive constant and linear intervals (top) and non-adaptive piecewise linear (bottom). Only 150~data points out of 600  are shown. Both segmentations have the same model complexity, but we simultaneously reduced the fit and the leave-one-out cross-validation error with the adaptive model (top).}
\end{figure} 

Typically, in a time series segmentation, a single model is applied to all intervals. 
For example, all intervals are assumed to behave in a quasi-constant or quasi-linear manner.
However, mixing different models and, in particular, constant and linear intervals can
have two immediate benefits. Firstly,
some applications need a qualitative description of each interval~\cite{1007574} indicated
by change of model: is 
the temperature rising, dropping or is it stable? In an ECG, we need to identify the flat interval between each cardiac pulses.
Secondly, as we will show, it can  reduce the fit error without increasing the cross-validation error.
Intuitively, a piecewise model tells when the data is increasing, and at what rate, and \textit{vice
versa}. While most time series have clearly identifiable linear trends some of the time,
this is not true over all time intervals. Therefore, the piecewise linear model
 locally overfits the data by computing meaningless slopes (see  Fig.~\ref{sunwplot}).  

Global overfitting has been addressed by limiting the number of 
 regressors~\cite{vasko2002}, but this carries the implicit assumption that time
 series are somewhat stationary~\cite{GaetanMonari06012002}.
Some frameworks~\cite{1007574} qualify the intervals where the slope is not significant
as being ``constant'' while others look for constant intervals within upward or downward intervals~\cite{YLBIJCAI05}.


Piecewise linear segmentation is ubiquitous and was one of the first
applications of dynamic programming~\cite{bellman1969}. We argue that many
applications can benefit from replacing it with a mixed model (piecewise linear and constant).
When identifying
constant intervals \textit{a posteriori} from a piecewise linear model, we risk misidentifying  some patterns including 
``stair cases'' or ``steps'' (see Fig.~\ref{sunwplot}).
A contribution of this paper is experimental evidence that we reduce fit without sacrificing the cross-validation error or running time for a given model complexity by using adaptive algorithms where 
some  intervals have a constant model whereas others have a linear model.
The new heuristic
we propose is neither more difficult to implement nor more expensive computationally. Our experiments
include white noise and random walks as well as ECGs and stock market prices. We also 
compare against the dynamic programming (optimal) solution which we show can be computed
in time $O(n^2 k)$.

Performance-wise, common heuristics (piecewise linear or constant)
have been reported to require quadratic time~\cite{657889}. 
We want fast linear time algorithms. When the number of desired segments is
small, top-down heuristics might be the only sensible option. We show
that if we allow the one-time
linear time computation of a buffer, adaptive (and non-adaptive) top-down heuristics  run in linear time ($O(n)$).

Data mining requires  high scalability over very large data sets.
Implementation issues, including numerical
stability, must be considered. In this paper,
we present algorithms that can process millions of data points in minutes, not hours.

\section{Related Work}

Table~\ref{mycomp} summarizes the various common heuristics and algorithms used to solve
the segmentation problem with polynomial models while minimizing the Euclidean ($l_2$) error. The top-down heuristics are
described in section~\ref{sect-topdown}. 
When  the number
of data points $n$ is much larger than the number of segments $k$ ($n \gg k$),  the
top-down heuristics is particularly competitive. Terzi and Tsaparas~\cite{DBLP:conf/sdm/TerziT06} 
achieved a complexity of $O(n^{4/3} k^{5/3} )$  for the piecewise constant model by running $(n/k)^{2/3}$ dynamic programming
routines and using weighted segmentations.
The original dynamic programming solution proposed by Bellman~\cite{bellman1969} ran in time $O(n^3 k)$,
and while it is known that a $O(n^2 k)$-time implementation is possible for piecewise constant
segmentation~\cite{DBLP:conf/sdm/TerziT06}, we will show in this paper that the same reduced complexity applies for piecewise linear
and mixed models segmentations as well.

\begin{table}
\caption{\label{mycomp}Complexity of various segmentation algorithms using 
polynomial models with $k$~segments and $n$~data points, including the
exact solution by dynamic programming.}
\begin{center}
\begin{tabular}{cc}
Algorithm & Complexity \\ \hline
Dynamic Programming & $O(n^2 k)$ \\ 
Top-Down & $O(n k)$ \\ 
Bottom-Up & $O(n \log n )$~\cite{palpanas04online} or $O(n^2/k)$~\cite{657889} \\
Sliding Windows & $O(n)$~\cite{palpanas04online} \\
\hline
\end{tabular}
\end{center}
\end{table}

Except for Pednault who mixed linear and quadratic segments~\cite{Pednault1991},
we know of no other attempt to segment
time series using polynomials of variable degrees in the data mining and knowledge discovery
literature though there is related work in the spline and statistical literature~\cite{Friedman1991,Lindstrom99,Zuo97} and machine learning literature~\cite{allen1974rbv,atkeson1997lwl,birattari1999llm}.
The introduction of ``flat'' intervals in a segmentation model 
has been addressed previously in the context of quasi-monotonic 
segmentation~\cite{YLBIJCAI05} by identifying flat subintervals within increasing or decreasing
intervals, but without concern for the cross-validation error.

While we focus on segmentation, there are many methods available for fitting models to continuous variables, such as a regression, regression/decision trees, Neural Networks~\cite{Hastie2001}, Wavelets~\cite{Donoho1994}, Adaptive Multivariate Splines~\cite{Friedman1991}, Free-Knot Splines~\cite{Lindstrom99}, Hybrid Adaptive Splines~\cite{Zuo97}, etc.

\section{Complexity Model}

Our complexity model  is purposely simple. The model complexity of a segmentation is
the sum of the number of regressors over each interval: a constant 
interval has a cost of 1, a linear interval a cost of 2 and so on. In other words, 
 a linear interval is as complex as two constant intervals.
Conceptually,
regressors are real numbers whereas all other parameters describing the model
only require a small number of bits.

\begin{figure}[h]
	\center\includegraphics[width=8cm,angle=0]{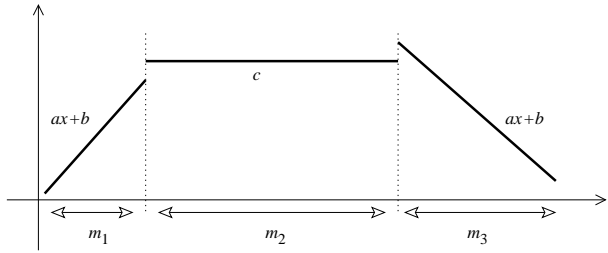}
\caption{\label{multidegreemodel}To describe an adaptive segmentation, you need
the length and regressors of each interval.}
\end{figure}

In our implementation, each regressor counted uses 64~bits 
(``double'' data type in modern
C or Java).
There are two types of  hidden parameters which we discard (see Fig.~\ref{multidegreemodel}):
the width or location of the intervals and the number of regressors per interval. 
The number of regressors per interval is only a few bits and is not
 significant in all cases. The width of the intervals in number of data points
can be represented using $\kappa\lceil \log m\rceil $~bits where $m$ is the maximum length of a
interval and $\kappa$ is the number of intervals: in the experimental cases we considered, $\lceil\log m\rceil \leq 8$  which is small compared to 64, the number of bits used to store each regressor counted.
We should also consider that slopes typically need to be stored using more accuracy (bits)
than constant values. This last consideration is not merely theoretical since a 32~bits
implementation of our algorithms is possible for the piecewise constant model whereas,
in practice, we require 64~bits for the piecewise linear model (see proposition~\ref{prop:fastpolyfit} and
discussion that follows).
Experimentally, the piecewise linear model can significantly outperform (by $\approx 50\%$) the 
piecewise constant model in accuracy (see  Fig.~\ref{ecgplot}) and vice versa.
For the rest of this paper, we take the fairness of our complexity model as an axiom. 

The desired total number of regressors depends on domain knowledge and the application:
when processing ECG data, whether we want to have two intervals per cardiac pulse or 20 intervals depends on whether we are satisfied with the mere identification of the general location of the pulses or whether we desire a finer analysis. In some instances, the user has no guiding
principles or domain knowledge from which to choose the number of intervals and a model
 selection algorithm is needed.
Common model selection approaches such as Bayesian Information Criterion~(BIC), 
Minimum Description Length~(MDL) and Akaike Information Criterion~(AIC) suffer because the possible
model complexity $p$ is large in a segmentation problem ($p=n$)~\cite{Burnham2004}. More conservative model selection approaches such as Risk Inflation Criterion~\cite{Foster1994} or Shrinkage~\cite{Donoho1994} do not directly apply because they assume wavelet-like regressors. Cross-validation~\cite{Friedl2002}, generalized cross-validation~\cite{343930},  and leave-one-out cross-validation~\cite{Tsuda2000} methods are too expensive.
However, stepwise regression analysis~\cite{363580} techniques such as permutation tests (``pete'') are far more practical~\cite{vasko2002}. In this paper, 
we assume that the model complexity is known either as an input from the user 
or through model selection.

\section{Time Series, Segmentation Error and Leave-One-Out}

Time series are sequences of data points  $(x_0,y_0),\ldots,(x_{n-1},y_{n-1}) $  
where the $x$ values, the ``time'' values, are sorted: $x_i>x_{i-1}$. In this paper, both
the $x$ and $y$ values are real numbers.
We define a segmentation as a sorted set of segmentation indexes $z_0, \ldots, z_{\kappa}$ such
that $z_0=0$ and $z_{\kappa}=n$. The segmentation points divide the time series into intervals $S_1,\ldots, S_\kappa$ defined by the segmentation indexes as $S_j=\{ (x_i,y_i)| z_{j-1} \leq i < z_{j}\}$ . Additionally, each interval $S_1,\ldots, S_\kappa$ has a model (constant, linear, upward monotonic, and so on).

In this paper, the segmentation error is computed from $\sum_{j=1}^\kappa Q(S_j)$
where the function $Q$ is the square of the $l_2$ regression error.
Formally, $
Q(S_j) = \min_p \sum_{r=z_{j-1}}^{z_{j}-1} (p(x_r)-y_r)^2
          $
 where the minimum is over the polynomials $p$ of a given degree.
For example, if the interval $S_j$ is said to be
constant, then $Q(S_j)= \sum_{z_j \leq l\leq z_{j+1}} (y_l-\bar y)^2$ 
where $\bar y$ is the average, $\bar y= \sum_{z_{j-1} \leq l< z_{j}}
\frac{y_l}{z_{j+1}-z_j}.$
Similarly, if the interval has a linear model, then $p(x)$ is chosen to be the linear
polynomial $p(x)=ax+b$ where $a$ and $b$ are found by regression.
The segmentation error can be generalized to other norms, such as the maximum-error ($l_\infty$)
norm~\cite{YLBIJCAI05,YLBICDMO05} by replacing the $\sum$ operators by $\max$ operators.

When reporting experimental error, we use the $l_2$ error $\sqrt {\sum_{j=1}^\kappa Q(S_j)}$.
We only compare time series having a fixed number of data points, but otherwise, the mean square error
should be used: $\sqrt {\frac {\sum_{j=1}^\kappa Q(S_j)}{n}}$.

If the data follows the model over each interval, then the error is zero. For example, given the time series
 $(0,0), (1,0), (2,0), (3,1),(4,2),$ we get no error when choosing the segmentation indexes $z_0=0,z_1=2,z_2=5$
 with a constant model over the index interval $[0,2)$ and a linear model over the index interval $[2,5)$.
 However, the choice of the best segmentation is not unique: we also get no error by choosing the alternative segmentation indexes $z_0=0,z_1=3,z_2=5.$

There are two types of segmentation problem:
\squishlist
 \item given a bound on the model complexity, find the segmentation minimizing the segmentation error;
 \item given a bound on the segmentation error, find a segmentation minimizing the model complexity.
\squishend
If we can solve efficiently and incrementally one 
problem type, then the second problem type is indirectly solved. 
Because it is intuitively easier to suggest a reasonable bound on the
model complexity, we focus on the first problem type.

For applications such as queries by humming~\cite{zhu2003qht},
it is useful to bound the distance between two time series  using only 
the segmentation data (segmentation points and polynomials over each interval).
Let $\Vert \cdot \Vert$ be any norm in a Banach space, including the Euclidean distance.
 Given time series $y,y'$, let $s(y), s(y')$ be the piecewise polynomial approximations
 corresponding  to the segmentation, then by the triangle inequality
\komment{\begin{eqnarray*}
& \Vert s(y)-s(y') \Vert -  \Vert s(y)-y \Vert-\Vert s(y')-y' \Vert\\
& \leq \Vert y-y' \Vert \leq \\
& \Vert s(y)-s(y') \Vert  + \Vert s(y)-y\Vert +\Vert s(y')-y' \Vert  .
\end{eqnarray*}}
$\Vert s(y)-s(y') \Vert -  \Vert s(y)-y \Vert-\Vert s(y')-y' \Vert
 \leq \Vert y-y' \Vert \leq 
 \Vert s(y)-s(y') \Vert  + \Vert s(y)-y\Vert +\Vert s(y')-y' \Vert  .$
Hence, as long as the approximation errors are small,  $\Vert s(y)-y\Vert <\epsilon$
and $\Vert s(y')-y'\Vert <\epsilon$, then we have that  nearby segmentations 
imply nearby time series ($\Vert s(y)-s(y') \Vert  <\epsilon  \Rightarrow \Vert y-y' \Vert < 3\epsilon$)
and nearby time series imply nearby segmentations ($\Vert y-y' \Vert  <\epsilon  \Rightarrow \Vert s(y)-s(y') \Vert < 3\epsilon$).
This result is entirely general.

Minimizing the fit error is important.
On the one hand, if we do not assume that the approximation errors are small, it is possible 
for the segmentation data to be identical $\Vert s(y)-s(y') \Vert=0$
while the distance between the time series, $\Vert y-y' \Vert$, is large,
causing false positives when identifying patterns from the segmentation data. 
For example, the sequences $100,-100$ and $-100,100$
can be both approximated by the same flat model ($0,0$), yet
they are far apart. On the other hand, if the fit error is large, similar time series
can have different segmentations, thus causing false negatives.
For example, the sequences $-100,100, -100.1$ and $-100.1, 100,-100$
have the piecewise flat model approximations $0,0,-100.1$ and $-100.1, 0,0$
respectively.

Beside the  data fit error, another interesting form
of error is obtained by cross-validation: divide your data points into two sets (training and test),
and measure how well your model, as fitted over the training set, predicts the test set. 
We predict a missing data point $(x_i,y_i)$  by first determining 
the interval $[z_{j-1},z_{j})$ corresponding to the data point ($x_{z_{j_1}}<x_i<x_{z_j}$) and then we compute
$p(x_i)$ where $p$ is the regression polynomial over $S_j$. The error is $\vert p(x_i)-y_i \vert $.
We opt for the leave-one-out cross-validation where the test set is a single data point and
the training set is the remainder. We repeat
the cross-validation over all possible missing data points, except for the first and last data
point in the time series, and compute the mean square error. If computing the segmentation takes linear time, then 
computing the leave-one-out error in this manner takes quadratic time, which is prohibitive for 
long time series. 

Naturally, beyond the cross-validation and fit errors, a segmentation should provide
the models required by the application. A rule-based system might require to know where
the data does not significantly increase or decrease and if flat intervals have not been 
labelled, such queries are hard to support elegantly.

\section{Polynomial Fitting in Constant Time}
\label{polyfit}

The naive fit error computation over a given interval
takes linear time $O(n)$: solve for the polynomial $p$ and then
compute $\sum_i (y_i-p(x_i))^2$. This has lead other authors to
conclude that top-down segmentation algorithm such as Douglas-Peucker's
require quadratic time~\cite{657889} while we will show they can run
in linear time. 
 To segment a time 
series into quasi-polynomial intervals in optimal time, 
we must compute fit errors in constant time ($O(1)$).

\begin{proposition}\label{prop:fastpolyfit}
Given a time series $\{(x_i,y_i)\}_{i=1,\ldots,n}$, if 
we allow the one-time $O(n)$ computation of a prefix buffer, finding the best
polynomial fitting the data over the interval $[x_p,x_q]$ is $O(1)$. This is true whether we use the Euclidean distance ($l_2$) or higher order
norms ($l_r$ for $\infty>r>2$).
\end{proposition}

\begin{proof}
We prove the result using the Euclidean ($l_2$) norm, the proof is similar for
higher order norms. 

We begin by showing that polynomial regression can be reduced to a matrix inversion
problem. Given a polynomial
$\sum_{j=0}^{N-1} a_j x^j$, the square of the Euclidean error is  $\sum_{i=p}^{q}
(y_i - \sum_{j=0}^{N-1} a_j x_i^j)^2$.  Setting the derivative with
respect to $a_l$ to zero for $l=0,\ldots,N-1$, generates
a system of $N$ equations and $N$ unknowns, $\sum_{j=0}^{N-1} a_j \sum_{i=p}^{q} x_i^{j+l}= \sum_{i=p}^{q} y_i x_i^l $
where $l=0,\ldots,N-1$.  On the right-hand-side, we have a $N$
dimensional vector ($V_l=\sum_{i=p}^{q} y_i x_i^l$) whereas on the left-hand-side,
we have the $N \times N$  T\oe{}plitz  matrix
$A_{l,i}=\sum_{i=p}^{q} x_i^{i+l}$ multiplied by the  coefficients of the polynomial
($a_0,\ldots,a_{N-1}$). That is, we have the matrix-vector equation
$\sum_{i=0}^{N-1}A_{l,i} a_i = V_l$.

As long as $N\geq q-p$,
the matrix $A$ is invertible. When
$N < q-p$, the solution is given by setting $N=q-p$ and letting $a_i=0$ 
for $i>q-p$. 
Overall, when $N$ is bounded \textit{a priori} by a small integer, 
no expensive numerical analysis is needed. Only computing the matrix $A$
and the vector $V$ is potentially expensive because they involve summations
over a large number of terms.

Once the coefficients $a_0, \ldots, a_{N-1}$ are known, we compute the
fit error  using the formula:
\begin{eqnarray*}
 \sum_{i=p}^{q}
\left ( \sum_{j=0}^{N-1} a_j x_i^j - y_i  \right )^2 
& =&
  \sum_{j=0}^{N-1} \sum_{l=0}^{N-1} a_j a_l  \sum_{i=p}^{q} x_i^{j+l} \\
& -&  2 \sum_{j=0}^{N-1} a_j \sum_{i=p}^{q} x_i^{j} y_i +  \sum_{i=p}^{q} y_i^2.
\end{eqnarray*} 
Again, only the summations are potentially expensive.

 Hence, computing the best polynomial fitting some data points over
a specific range and computing the corresponding fit error in constant time 
is equivalent to computing range sums of the form 
$\sum_{i=p}^{q} x_i^{i} y_i^{l}$
in constant time for $0\leq i,l\leq 2N$. To do so, simply compute once all
prefix sums
$P^{j,l}_q = \sum_{i=0}^{q} x_i^{j} y_i^{l}$
and then use their subtractions to compute range queries
$\sum_{i=p}^{q} x_i^{j} y_i^{l} = P^{j,l}_q-P^{j,l}_{p-1}.$
\end{proof}

Prefix sums speed up the computation of the range sums (making them constant time)
at the expense of update time and storage: if one of the data point changes, we may have
to recompute the entire prefix sum. 
 More scalable algorithms are possible if the time series are dynamic~\cite{LemireCASCON2002}.  Computing the needed prefix sums is only  done
once in linear time and requires $(N^2 +N +1) n $ units of  storage ($6 n$ units when $N=2$).
For most practical purposes, we argue that
we will soon have infinite storage so that trading storage for
speed is a good choice.
It is also possible to use
less storage~\cite{Ola2004}.

When using floating point values, 
the prefix sum approach causes a  loss in numerical accuracy 
which becomes significant
if $x$ or $y$ values grow large and $N>2$ (see Table~\ref{constanttimefitting}).
When $N=1$ (constant polynomials), 32~bits floating point numbers are sufficient,
but for $N\geq 2$, 64~bits is required.
In this paper, we are not interested in higher order polynomials 
and choosing $N=2$ is sufficient.

\begin{table}[tb]
\caption{\label{constanttimefitting}Accuracy of the polynomial fitting in constant time
using 32~bits and 64~bits floating point numbers respectively. We give
the worse percentage of error over 1000~runs using uniformly distributed white noise ($n=200$).
The domain ranges from $x=0$ to $x=199$ and we compute the fit error over
the interval $[180,185)$.}
\begin{center}
\begin{tabular}{l|cc}
\hline \hline
 & 32~bits& 64~bits    \\ \hline
$N=1$ ($y=b$) & $7\times 10^{-3}$\% & $1\times 10^{-11}$\% \\
$N=2$ ($y=ax+b$)& $5$\% & $6\times 10^{-9}$\%  \\
$N=3$ ($y=ax^2+bx+c$)& $240$\% & $3\times 10^{-3}$\%\\
\end{tabular}
\end{center}
\end{table}

\section{Optimal Adaptive Segmentation}

An algorithm is optimal, if it can find a
segmentation with minimal error  given a model complexity $k$.
Since we can compute best fit error in constant time for arbitrary polynomials,
a dynamic programming algorithm  computes
the optimal adaptive segmentation in time $O(n^2 N k)$ where $N$ is the upper
bound on the polynomial degrees. Unfortunately, if $N\geq 2$, this result 
does not hold in practice with 32~bits floating point numbers (see Table~\ref{constanttimefitting}).

We improve over the classical
approach~\cite{bellman1969} because we allow the polynomial degree of each interval
to vary. 
In the tradition of dynamic programming~\cite[pages 261--265]{kleinberg2006ad}, in a first stage, we compute the optimal cost matrix ($R$):
$R_{r,p}$ is the minimal segmentation cost of the time interval $[x_0,x_p)$ using a model complexity of
 $r$. If $E(p,q,d)$ is the fit error of a polynomial of degree 
$d$ over the time interval $[x_p,x_q)$, computable in time $O(1)$ by proposition~\ref{prop:fastpolyfit}, then 
\[R_{r,q}=\min_{0 \leq p \leq q,0 \leq d < N} R_{r-1-d,p}+E(p,q,d)\]
with the convention that $R_{r-1-d,p}$ is infinite when $r-1-d <0$ except for $R_{-1,0}=0$.
Because computing $R_{r,q}$ only requires knowledge of the prior rows, $R_{r',\cdot}$ for $r'<r$, we can compute $R$
 row-by-row starting with the first row (see Algorithm~\ref{algo:dynprog}).
Once we have computed the $r \times n+1$ matrix, we reconstruct the optimal
solution with a simple $O(k)$
algorithm (see Algorithm~\ref{algo:dynprog2}) using matrices $D$ and $P$ storing
respectively the best segmentation points and the best degrees.



\begin{algorithm}
 \begin{algorithmic}[1]
  \STATE \textbf{INPUT:} Time Series $(x_i,y_i)$ of length $n$
  \STATE \textbf{INPUT:} Model Complexity $k$ and maximum degree $N$ ($N=2 \Rightarrow$ constant and linear)
  \STATE \textbf{INPUT:} Function $E(p,q,d)$ computing fit error with poly. of degree $d$ in range $[x_p,x_q)$ (constant time)
  \STATE $R,D,P \leftarrow$ $k\times n+1$ matrices (initialized at 0) 
  \FOR{$r \in \{0,\ldots,k-1\}$}
  \STATE \COMMENT{$r$ scans the rows of the matrices}
    \FOR{$q \in \{0,\ldots,n\}$}
    \STATE \COMMENT{$q$ scans the columns of the matrices}
    
      \STATE Find a minimum 
      of $R_{r-1-d,p}+E(p,q,d)$ and store its value in $R_{r,q}$, and the corresponding $d,p$
      tuple in $D_{r,q},P_{r,q}$ for $0\leq d \leq\min(r+1,N)$ and  $0 \leq p \leq q+1$
      with the convention that $R$ is $\infty$ on negative rows except for $R_{-1,0}=0$.
      \ENDFOR
    \ENDFOR
  
   \STATE \textbf{RETURN} cost matrix $R$, degree matrix $D$, segmentation points matrix $P$
 \end{algorithmic}
\caption{\label{algo:dynprog}First part of dynamic programming algorithm for optimal adaptive segmentation
 of time series into intervals having degree $0,\ldots,N-1$. }
\end{algorithm}
\begin{algorithm}
\begin{algorithmic}[1]
\STATE \textbf{INPUT:} $k\times n+1$ matrices $R$, $D$, $P$ from dynamic programming algo.
\STATE $x \leftarrow n$
\STATE $s \leftarrow $ empty list
\WHILE{$r \geq 0$}
\STATE $p \leftarrow P_{r,x}$
\STATE $d \leftarrow D_{r,x}$
\STATE $r \leftarrow r-d+1$
\STATE append interval from $p$ to $x$ having degree $d$ to $s$
\STATE $x \leftarrow p$
\ENDWHILE
\STATE \textbf{RETURN} optimal segmentation  $s$
 \end{algorithmic}
\caption{\label{algo:dynprog2}Second part  of dynamic programming algorithm for optimal adaptive segmentation. }
\end{algorithm}

\section{Piecewise Linear or Constant Top-Down Heuristics}

\label{sect-topdown}

Computing optimal segmentations with dynamic programming 
is  $\Omega (n^2)$  which is not practical when the size of the time series
is  large. Many efficient online approximate algorithms are based on greedy strategies. 
Unfortunately, for small model complexities, popular  heuristics run in quadratic time ($O(n^2)$)~\cite{657889}.
Nevertheless, when the desired model complexity is small, 
a particularly effective heuristic is  the top-down  segmentation which 
proceeds as follows: starting with a simple segmentation, we further segment
 the worst interval, and so on, until we exhaust the budget. Keogh \textit{et al.}~\cite{657889} state that
this algorithm  has been independently discovered
in the seventies and is known by several
name: Douglas-Peucker algorithm, Ramers algorithm, or Iterative End-Points Fits.  
In theory, Algorithm~\ref{algo:topdown} computes the top-down segmentation, using polynomial regression
of any degree, in time $O(kn)$ where $k$ is the model complexity, 
by using fit error computation in constant time. In practice, our implementation only 
works reliably for $d=0$ or $d=1$ using 64~bits floating point
numbers.
The piecewise constant ($d=0$) and  piecewise linear ($d=1$) cases are referred to as the ``top-down constant'' and ``top-down linear'' heuristics respectively.

\begin{algorithm}
 \begin{algorithmic}
\STATE \textbf{INPUT:} Time Series $(x_i,y_i)$ of length $n$
\STATE \textbf{INPUT:} Polynomial degree $d$ ($d=0$, $d=1$, etc.) and model complexity $k$
\STATE \textbf{INPUT:} Function $E(p,q)$ computing fit error with poly. in range $[x_p,x_q)$
\STATE $S$ empty list
 \STATE $S\leftarrow (0,n, E(0,n))$
\STATE $b \leftarrow k-d$
 \WHILE{ $b-d\geq 0$ }
\STATE find tuple $(i,j,\epsilon)$ in $S$ with maximum last entry 
\STATE find minimum of $E(i,l)+E(l,j)$ for $l=i+1,\ldots,j$
\STATE remove tuple $(i,j,\epsilon)$ from $S$
\STATE insert tuples $(i,l,E(i,l))$ and $(l,j,E(l,j))$ in $S$
\STATE $b \leftarrow b-d$
 \ENDWHILE
\STATE $S$ contains the segmentation
 \end{algorithmic}
\caption{\label{algo:topdown}Top-Down Heuristic.}
\end{algorithm}

\section{Adaptive Top-Down Segmentation}

\begin{figure}
	\center\includegraphics[width=\standardlength,angle=0]{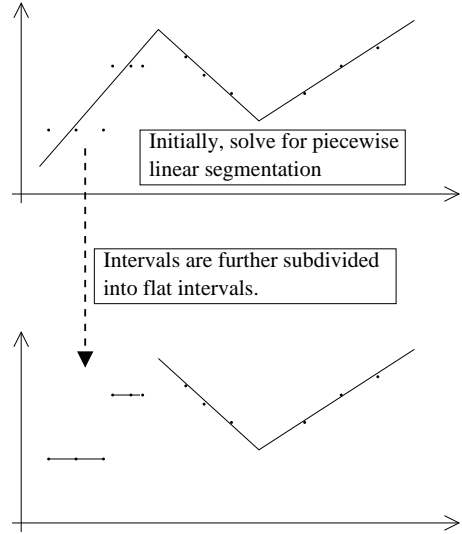}
\caption{\label{adaptiveheuristic}Adaptive Top-Down Segmentation: initially, we compute a
piecewise linear segmentation, then we further subdivide some intervals
into constant intervals.}
\end{figure}

Our linear time adaptive segmentation heuristic is based on the
observation that a linear interval can be
replaced  by two constant intervals without model complexity increase.
After applying the top-down linear heuristic from the previous section 
(see Algorithm~\ref{algo:topdown}), we  optimally
subdivide each interval once with intervals having fewer regressors (such as
constant) but the same total model complexity. The computational complexity is
the same, $O((k+1)n)$.  The result is Algorithm~\ref{algo:adaptivetopdown} as illustrated by
Fig.~\ref{adaptiveheuristic}. In practice, we first apply the top-down linear heuristic
and then we seek to split the linear intervals into two constant intervals.

Because the algorithm only splits an interval
if the fit error can be reduced, it is guaranteed not to degrade the fit error. However,
improving the
fit error is not, in itself, desirable unless we can also ensure we do not increase the cross-validation error.

An alternative strategy is to proceed from the top-down constant
heuristic and try to merge constant intervals into linear intervals. We chose not to report
our experiments with this alternative since, over our data sets, it gives worse results
and is slower than all other heuristics.

\begin{algorithm}
 \begin{algorithmic}
\STATE \textbf{INPUT:} Time Series $(x_i,y_i)$ of length $n$
\STATE \textbf{INPUT:} Bound on Polynomial degree $N$ and model complexity $k$
\STATE \textbf{INPUT:} Function $E(p,q,d)$ computing fit error with poly. in range $[x_p,x_q)$
\STATE $S$ empty list
\STATE $d \leftarrow N-1$
 \STATE $S\leftarrow (0,n,d, E(0,n,d))$
\STATE $b \leftarrow k-d$
 \WHILE{ $b-d\geq 0$ }
\STATE find tuple $(i,j,d,\epsilon)$ in $S$ with maximum last entry 
\STATE find minimum of $E(i,l,d)+E(l,j,d)$ for $l=i+1,\ldots,j$
\STATE remove tuple $(i,j,\epsilon)$ from $S$
\STATE insert tuples $(i,l,d,E(i,l,d))$ and $(l,j,d,E(l,j,d))$ in $S$
\STATE $b \leftarrow b-d$
 \ENDWHILE
\FOR {tuple $(i,j,q,\epsilon)$ in $S$}
\STATE find minimum $m$ of $E(i,l,d')+E(l,j,q-d'-1)$ for $l=i+1,\ldots,j$ and $0 \leq d' \leq q-1$
\IF{ $m < \epsilon$ } 
\STATE remove tuple $(i,j,q,\epsilon)$ from $S$
\STATE insert tuples $(i,l,d',E(i,l,d'))$ and $(l,j,q-d'-1,E(l,j,q-d'-1))$ in $S$
\ENDIF
\ENDFOR
\STATE $S$ contains the segmentation
 \end{algorithmic}
\caption{\label{algo:adaptivetopdown}Adaptive Top-Down Heuristic.}
\end{algorithm}

\section{Implementation and Testing}

Using a Linux platform, we implemented our algorithms in C++ using GNU GCC~3.4 and flag
``-O2''. Intervals are stored in an STL \textit{list} object.
Source code is available from the author. 
Experiments run on a PC with an AMD Athlon~64~(2~GHZ)~CPU and enough internal memory
so that no disk paging is observed. 

\begin{figure}
	\center\includegraphics[height=\standardlength,angle=270]{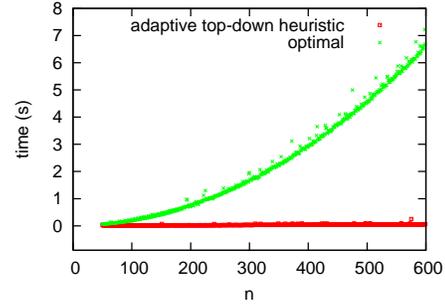}
\caption{\label{timings}Adaptive Segmentation Timings: Time in seconds versus
the number of data points using ECG data. We compare the optimal dynamic 
programming solution with the top-down heuristic ($k=20$).
}
\end{figure}

Using ECG data and various number of data points,
we benchmark the optimal algorithm, using dynamic programming, against the adaptive
top-down heuristic: Fig.~\ref{timings} demonstrates that the quadratic time nature
of the dynamic programming solution is quite prevalent 
($t\approx n^2/50000$~seconds) making it unusable in all but
toy cases, despite a C++ implementation: nearly a full year would be required to 
optimally segment a time series with 1~million data points! 
Even if we record only one data point every second for an hour, we still
generate 3,600~data points which would require about 4~minutes to segment!
Computing the leave-one-out error of a
quadratic time segmentation algorithm requires cubic time: to process
the numerous time series we chose for this paper, days of processing
are required. 

\begin{figure}
	\center\includegraphics[width=6.5cm,angle=0]{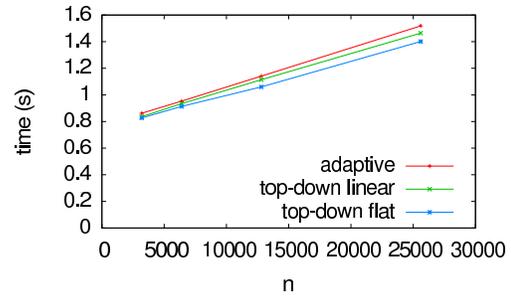}
\caption{\label{heutimings}Heuristics timings using random walk data: Time in seconds versus
the number of data points  ($k=20$). }
\end{figure}

We observed empirically that the timings
are not sensitive to the data source. 
The difference in execution time of the various heuristics is negligible (under 15\%):
our implementation of the adaptive heuristic is not significantly more expensive
than the top-down linear heuristic because its
additional step, where constant intervals are created out of linear ones, 
can be efficiently written as a simple sequential scan over the time series.
To verify the scalability, we generated random walk
time series of various length with fixed model complexity ($k=20$), see Fig.~\ref{heutimings}.

\section{Random Time Series and Segmentation}

\begin{figure}
  	\center\includegraphics[height=\standardlength,angle=270]{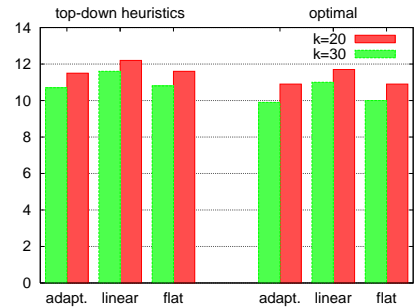}
 \caption{\label{randplot} Average Euclidean ($l_2$ norm) fit error over synthetic white noise data. }
  \end{figure}

Intuitively, adaptive algorithms over purely random data are wasteful. To verify
this intuition, we generated 10~sequences of Gaussian random noise ($n=200$): each data point
takes on a value according to a normal distribution ($\mu=0, \sigma=1$).
The average leave-one-out error is presented in Table~\ref{randomerror} (top) with model complexity $k=10,20,30$.
As expected, the adaptive heuristic shows a slightly worse cross-validation error. However,
this is compensated by a slightly better fit error (5\%) when compared
with the top-down linear heuristic (Fig.~\ref{randplot}). On this same figure,
we also compare the dynamic programming solution which shows a 10\% reduction in 
fit error for all three models (adaptive, linear, constant/flat), for twice the running time (Fig.~\ref{timings}).
The relative errors are not sensitive to the model complexity.

\begin{figure}
  	\center\includegraphics[height=\standardlength,angle=270]{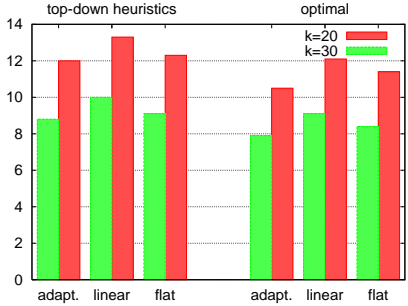}
 \caption{\label{walkplot} Average Euclidean ($l_2$ norm) fit error over synthetic random walk data. }
  \end{figure}

Many chaotic processes such as stock prices are sometimes described as random walk. Unlike
white noise, the value of a given data point depends on its neighbors. We generated 10~sequences of Gaussian random walks ($n=200$): starting at the value 0, each data point
 takes on the value $y_i=y_{i-1}+N(0,1)$ where $N(0,1)$ is a value from  a normal distribution ($\mu=0, \sigma=1$).
The results are presented in Table~\ref{randomerror} (bottom) and Fig.~\ref{walkplot}. The adaptive
algorithm improves the leave-one-out error (3\%--6\%) and the fit error ($\approx 13\%$) over the top-down
linear heuristic. Again, the optimal algorithms improve over the heuristics by approximately 10\%
and the model complexity does not change the relative errors.

\begin{table}
\caption{\label{randomerror}Leave-one-out cross-validation error for top-down heuristics on 10 sequences of  Gaussian white noise (top) and  random walks (bottom) ($n=200$) for various model complexities ($k$)}
\begin{center}\small
\begin{tabular}{ccccc}
\hline \hline
white noise & adaptive & linear  & constant  & linear/adaptive  \\ \hline
$k=10$ & 1.07 &  1.05 & \textbf{1.06} & 98\%  \\
$k=20$ & 1.21 &  1.17 & \textbf{1.14} & 97\%  \\
$k=30$ & 1.20 & 1.17 & \textbf{1.16} & 98\% \\
\end{tabular}
\begin{tabular}{ccccc}
\hline \hline
random walks & adaptive & linear  & constant  & linear/adaptive  \\ \hline
$k=10$  & \textbf{1.43 } & 1.51 & \textbf{1.43} & 106\% \\
$k=20$  & \textbf{1.16} & 1.21 & 1.19  & 104\% \\
$k=30$  & \textbf{1.03} & 1.06 & 1.06 & 103\% \\
\end{tabular}
\end{center}
\end{table}


\section{Stock Market Prices and Segmentation}

Creating, searching and identifying stock market patterns is sometimes done using
segmentation algorithms~\cite{1007574}. Keogh and Kasetty suggest~\cite{keogh2003nts}
that stock market data is indistinguishable from random walks. If so, the
good results
from the previous section should carry over. However, the random
walk model has been strongly rejected using variance estimators~\cite{lo1988smp}. Moreover, Sornette~\cite{sornette2002smc} claims stock markets are akin to physical systems 
and can be predicted. 

Many financial market experts look for patterns and trends
in the historical stock market prices, and this approach is called  
``technical analysis'' or ``charting''~\cite{507653,Balvers2000,Fama1988}.
If you take into account ``structural breaks,'' some stock market prices
have detectable locally stationary trends~\cite{Chaudhuri2003}.
Despite some controversy, technical analysis is a Data Mining topic~\cite{tse2002mai,fu2004fts,kamruzzaman2003sbm}.

\komment{
Technical analysis patterns are easier to understand when a scenario describes
 the stock's behavior, but one should not take such scenarios too seriously. 
Here are some common patterns that can be used to predict stock prices:
\begin{itemize} 
 \item A stock's price has grown several months according to a noticeable linear trend.
If the price ever grows above this linear trend, stock analysts may feel that the stock
is ``overvalued'' and sell, thus lowering the prices, and if the price stops growing linearly,
analysts may feel that the stock is suddenly undervalued and will buy, thus bringing the prices up.
In other words, clearly visible trends may be self-sustained (see Fig.~\ref{techanalysislinear}). The
same type of arguments can be given if the ``trend'' is for the stock prices to oscillate around
a fixed price.
\item Important fund managers instructs their staff to sell a given stock whenever
it grows beyond a fixed price. As the stock prices get past this limit, the sell orders
cause the prices to drop slightly, but as the prices drop some investors  buy again to 
benefit from the perceived bargain, thus causing the prices to get past the upper limit once more
causing new sell orders and a new drop in prices. After one or two iterations of this process,
few investors will be left to buy the stock as the prices drop and the price will fall.
This stock pattern is known as ``head and shoulders''~\cite{318357} (see Fig.~\ref{techanalysishands}).
\item Reversing the previous case and imagining that fund managers instruct their staff to 
buy a stock whenever it falls below a fixed price, we get the ``double bottom'' pattern.
\item As in the scenario we made up for the ``head and shoulders'' pattern,
imagine that a fund manager decides to sell
a given stock whenever it reaches a given price, but suppose it has a diminishing supply of
the stock to sell and other important investors think that the stock is worth more: 
eventually after some ups and downs, when the fund manager has exhausted its supply, the
stock value will raise again. This pattern is called the ``ascending triangle'' (see Fig.~\ref{techanalysisascendingtriangle}).
Naturally, the reverse pattern also exists (``descending triangle'').
\end{itemize}
}

\komment{
\begin{figure}
	\center\includegraphics[width=\standardlength,angle=0]{techanalysislinear.eps}
\caption{\label{techanalysislinear}Some trends in technical analysis are locally stable.}
\end{figure} 
\begin{figure}
	\center\includegraphics[width=\standardlength,angle=0]{techanalysishands.eps}
\caption{\label{techanalysishands}The ``head and shoulders'' pattern: it can be used
to predict a drop in the price of a stock.}
\end{figure} 
\begin{figure}
	\center\includegraphics[width=\standardlength,angle=0]{techanalysisascendingtriangle.eps}
\caption{\label{techanalysisascendingtriangle}The ``ascending triangle'' pattern: it can be used
to predict an increase in the price of a stock.}
\end{figure} 
}

\begin{figure}
  	\center\includegraphics[height=\standardlength,angle=270]{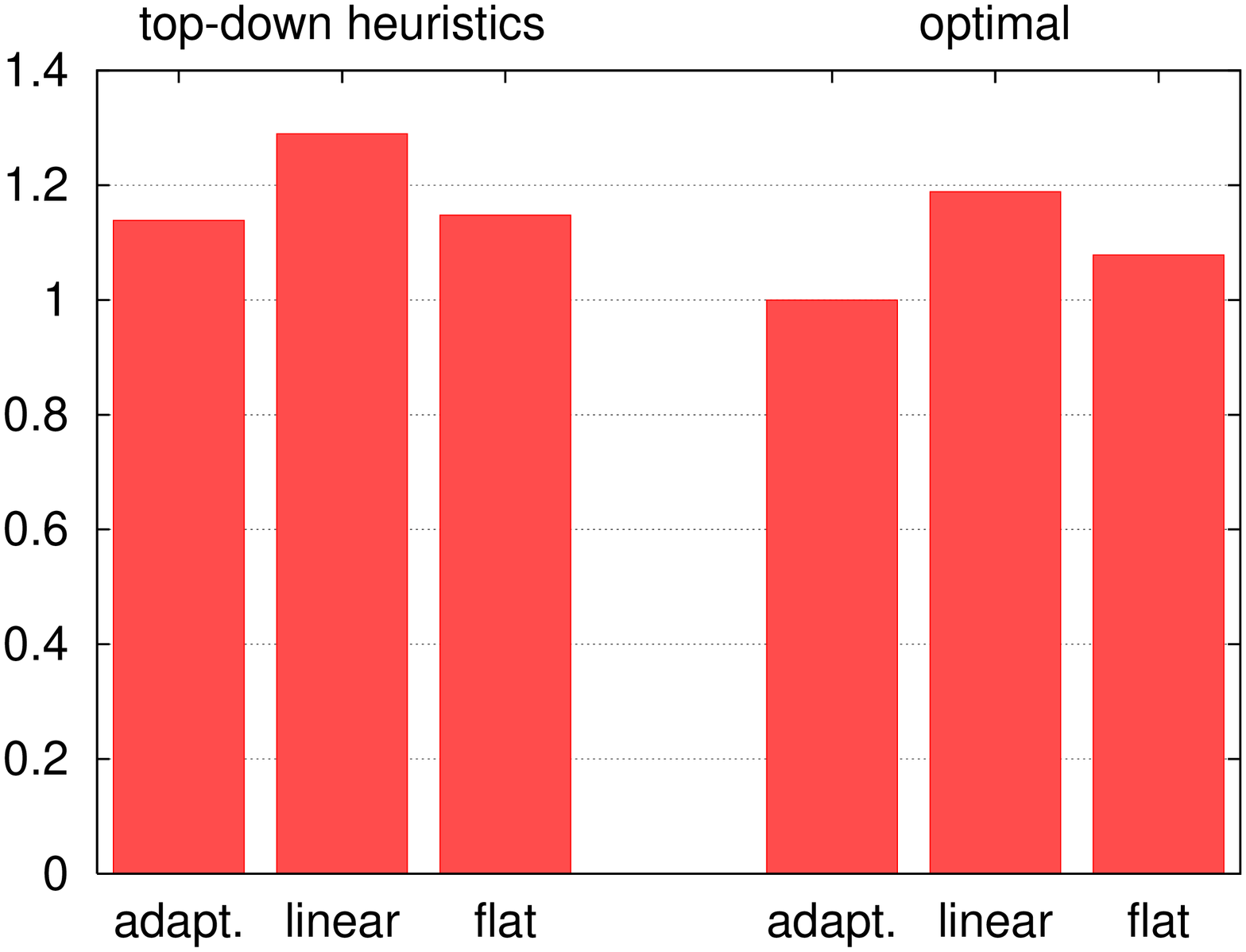}
 \caption{\label{stockplot} Average Euclidean ($l_2$ norm) fit error over 14~different stock prices, for each stock, the error was normalized ($1=$ optimal adaptive segmentation). The complexity is set at 20 ($k=20$) but relative results are not sensitive to the complexity.}
  \end{figure}

\begin{figure*}
	\center\includegraphics[height=0.3\textwidth,angle=270]{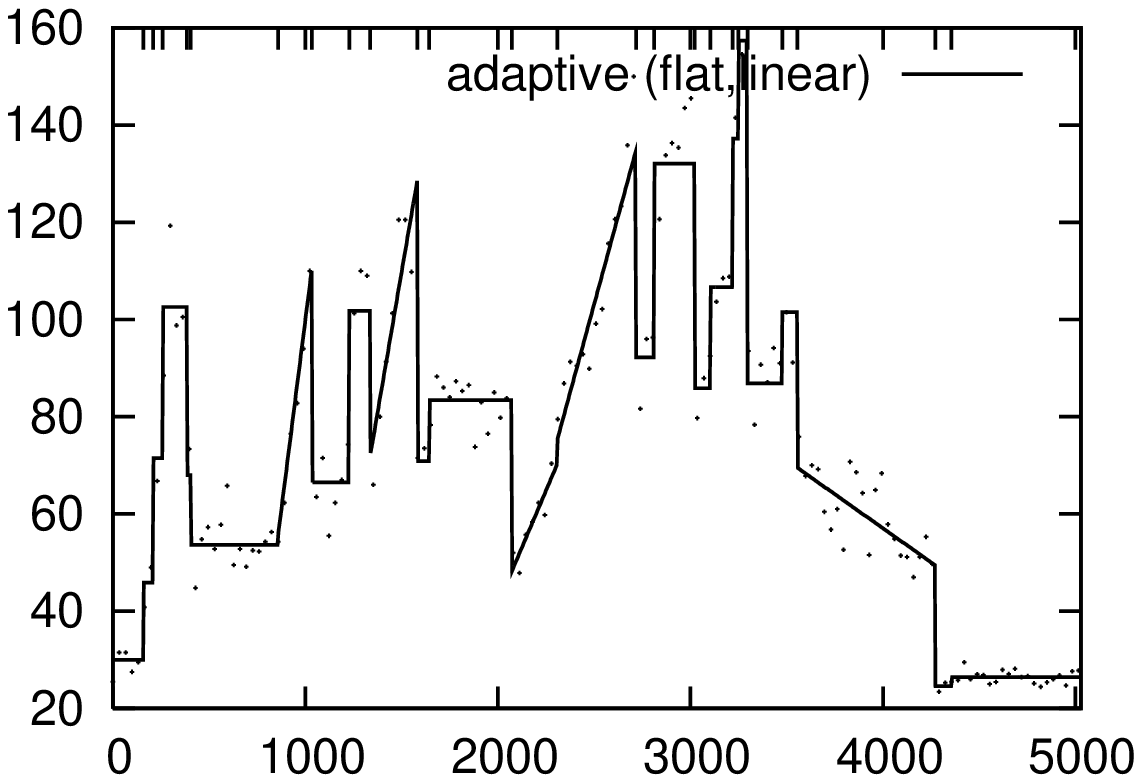}
\includegraphics[height=0.3\textwidth,angle=270]{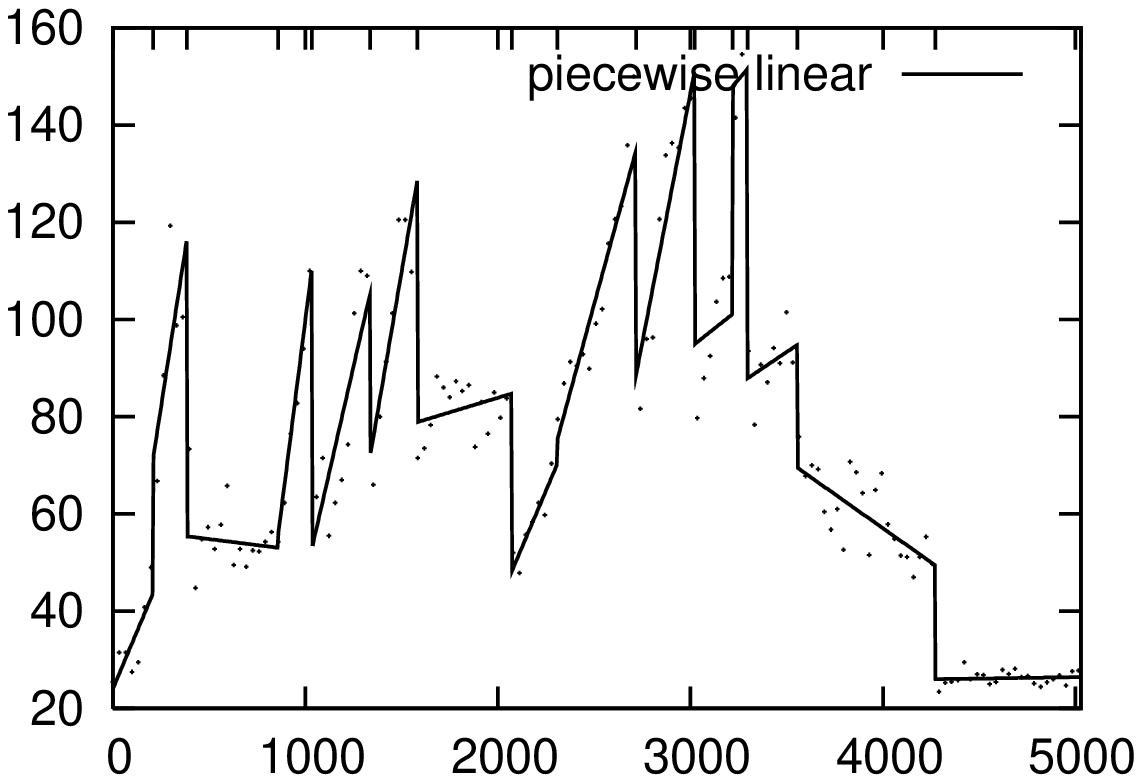}
\includegraphics[height=0.3\textwidth,angle=270]{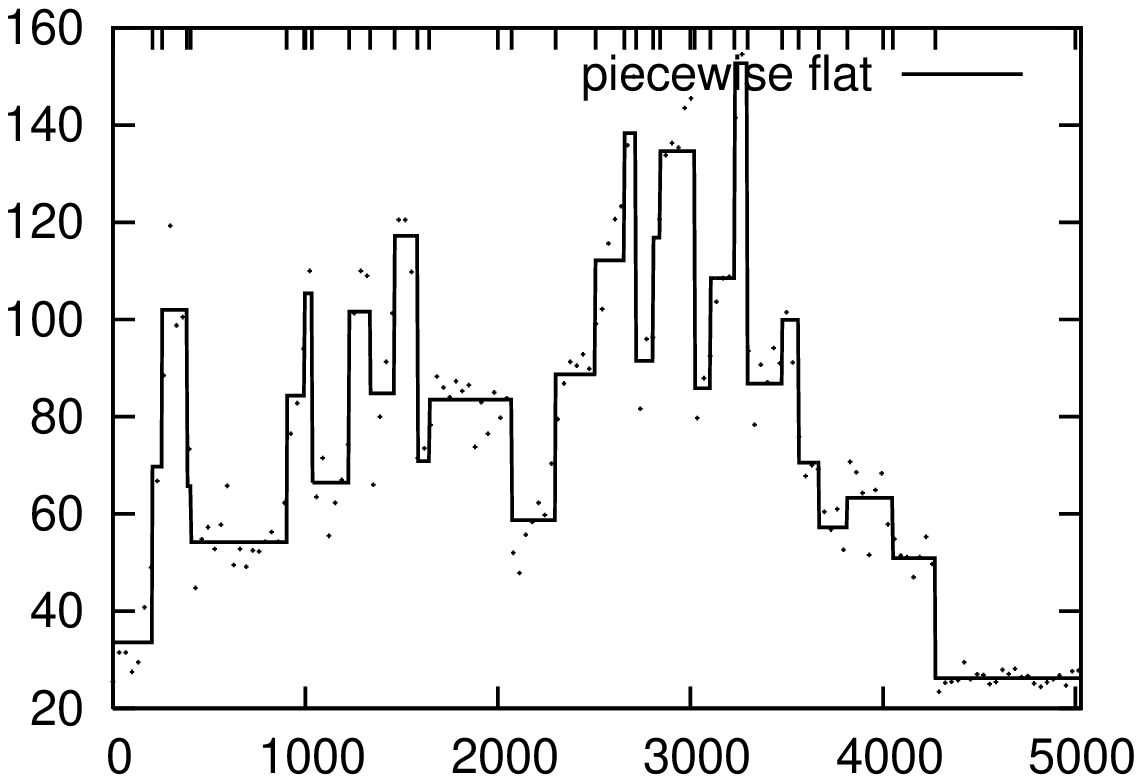}
\caption{\label{microsoft}Microsoft historical daily stock prices (close) segmented by the adaptive
top-down (top), linear top-down (middle), and constant top-down (bottom) heuristics.
\textbf{For clarity, only 150~data points out of 5,029  are shown.}}
\end{figure*}

We segmented daily stock prices from dozens of companies~\cite{yahoofinance}.
Ignoring stock splits, we pick the first 200~trading days of each stock or index. The model
complexity varies $k=10,20,30$ so that the number of intervals can range from 5 to 30.
We compute the segmentation error using 3~top-down~heuristics: adaptive, linear and constant (see Table~\ref{stockerror} for some of the result). As expected, the adaptive heuristic is more accurate
than the  top-down linear heuristic (the gains are between 4\% and 11\%). 
The  leave-one-out cross-validation error is improved with the adaptive heuristic when the model complexity
 is small.  The relative fit error is not sensitive to the model complexity. We observed similar results using other stocks. These results are consistent with our synthetic random walks results. Using all of the historical data available, 
we plot the 3~segmentations for Microsoft stock prices (see Fig.~\ref{microsoft}). The  line
is the regression polynomial over each interval and only  150~data points out of 5,029  are shown to avoid
clutter. Fig.~\ref{stockplot} shows the average fit error for all 3~segmentation models: in order to average the results, we first normalized the errors of each stock so that the optimal adaptive is 1.0. These results are
consistent with the random walk results (see Fig.~\ref{walkplot}) and they indicate that the adaptive model
is a better choice than the piecewise linear model.

\begin{table*}
\caption{\label{stockerror}Euclidean segmentation error ($l_2$ norm) and cross-validation error for \komment{various model complexities ($k$)}$k=10,20,30$: lower is better.
}
\begin{center}\small
\begin{tabular}{c|cccc|cccc}
\hline \hline
 &  & fit error  &  &  
&  & leave one out error  &   &   \\  %
  & adaptative & linear  & constant & linear/adaptive 
& adaptative & linear  & constant  & linear/adaptive  \\ \hline %
 Google & \textbf{79.3} & 87.6 & 88.1 & 110\%
 &  \textbf{2.6} & 2.7 & 2.8 & 104\%\\
 Sun Microsystems & 23.1 & 26.5 & \textbf{21.7} & 115\%
 &  1.5 & 1.5 & \textbf{1.4} & 100\%\\
 Microsoft & \textbf{14.4} & 15.5 & 15.5 & 108\%
 &  \textbf{1.1} & \textbf{1.1} & \textbf{1.1} & 100\%\\
 Adobe & 15.3 & 16.4 & \textbf{14.7} & 107\%
 &  \textbf{1.1} & \textbf{1.1} & 1.2 & 100\%\\
 ATI & 8.6 & 9.5 & \textbf{8.1} & 110\%
 &  \textbf{0.8} & 0.9 & \textbf{0.8} & 113\%\\
 Autodesk & \textbf{9.8} & 10.9 & 10.4 & 111\%
 &  \textbf{0.9} & \textbf{0.9} & \textbf{0.9} & 100\%\\
 Conexant & \textbf{32.6} & 34.4 & \textbf{32.6} & 106\%
 &  \textbf{1.7} & \textbf{1.7} & \textbf{1.7} & 100\%\\
 Hyperion & 39.0 & 41.4 & \textbf{38.9} & 106\%
 &  1.9 & 2.0 & \textbf{1.7} & 105\%\\
 Logitech & 6.7 & 8.0 & \textbf{6.3} & 119\%
 &  \textbf{0.7} & 0.8 & \textbf{0.7} & 114\%\\
 NVidia & 13.4 & 15.2 & \textbf{12.4} & 113\%
 &  \textbf{1.0} & 1.1 & \textbf{1.0} & 110\%\\
 Palm & 51.7 & 54.2 & \textbf{48.2} & 105\%
 &  \textbf{1.9} & 2.0 & 2.0 & 105\%\\
 RedHat & \textbf{125.2} & 147.3 & 145.3 & 118\%
 &  \textbf{3.7} & 3.9 & \textbf{3.7} & 105\%\\
 RSA & 17.1 & 19.3 & \textbf{15.1} & 113\%
 &  \textbf{1.3} & \textbf{1.3} & \textbf{1.3} & 100\%\\
 Sandisk & 13.6 & 15.6 & \textbf{12.1} & 115\%
 &  1.1 & 1.1 & \textbf{1.0} & 100\%\\ \hline
 \komment{\hline
 $k=20$ &  &   & 
&  &   &    \\ \hline}
Google & \textbf{52.1} & 59.1 & 52.2 & 113\%
 &  \textbf{2.3} & 2.4 & 2.4 & 104\%\\
 Sun Microsystems & 13.9 & 16.5 & \textbf{13.8} & 119\%
 &  \textbf{1.4} & \textbf{1.4} & \textbf{1.4} & 100\%\\
 Microsoft & \textbf{10.5} & 12.3 & 11.1 & 117\%
 &  \textbf{1.0} & 1.1 & \textbf{1.0} & 110\%\\
 Adobe & 8.5 & 9.4 & \textbf{8.3} & 111\%
 &  1.1 & \textbf{1.0} & 1.1 & 91\%\\
 ATI & 5.2 & 6.1 & \textbf{5.1} & 117\%
 &  \textbf{0.7} & \textbf{0.7} & \textbf{0.7} & 100\%\\
 Autodesk & 6.5 & 7.2 & \textbf{6.2} & 111\%
 &  \textbf{0.8} & \textbf{0.8} & \textbf{0.8} & 100\%\\
 Conexant & \textbf{21.0} & 22.3 & 21.8 & 106\%
 &  \textbf{1.4} & \textbf{1.4} & 1.5 & 100\%\\
 Hyperion & \textbf{26.0} & 29.6 & 27.7 & 114\%
 &  1.8 & 1.8 & \textbf{1.7} & 100\%\\
 Logitech & \textbf{4.2} & 4.9 & \textbf{4.2} & 117\%
 &  \textbf{0.7} & \textbf{0.7} & \textbf{0.7} & 100\%\\
 NVidia & \textbf{9.1} & 10.7 & \textbf{9.1} & 118\%
 &  \textbf{0.9} & 1.0 & 1.0 & 111\%\\
 Palm & 33.8 & 35.2 & \textbf{31.8} & 104\%
 &  1.9 & 1.9 & \textbf{1.8} & 100\%\\
 RedHat & \textbf{77.7} & 88.2 & 82.8 & 114\%
 &  3.6 & 3.6 & \textbf{3.5} & 100\%\\
 RSA & \textbf{9.8} & 10.6 & 10.9 & 108\%
 &  1.2 & \textbf{1.1} & 1.2 & 92\%\\
 Sandisk & 9.0 & 10.6 & \textbf{8.5} & 118\%
 &  1.0 & 1.0 & \textbf{0.9} & 100\%\\ \hline
\komment{\hline 
$k=30$ & &   &  & 
&  &   &  &    \\ \hline}
 Google & \textbf{37.3} & 42.7 & 39.5 & 114\%
 &  \textbf{2.2} & \textbf{2.2} & 2.3 & 100\%\\
 Sun Microsystems & 11.7 & 13.2 & \textbf{11.6} & 113\%
 &  \textbf{1.4} & \textbf{1.4} & \textbf{1.4} & 100\%\\
 Microsoft & \textbf{7.5} & 9.2 & 8.4 & 123\%
 &  \textbf{1.0} & \textbf{1.0} & \textbf{1.0} & 100\%\\
 Adobe & \textbf{6.2} & 6.8 & \textbf{6.2} & 110\%
 &  1.0 & \textbf{0.9} & 1.0 & 90\%\\
 ATI & \textbf{3.6} & 4.5 & 3.7 & 125\%
 &  \textbf{0.7} & \textbf{0.7} & \textbf{0.7} & 100\%\\
 Autodesk & \textbf{4.7} & 5.5 & 4.9 & 117\%
 &  0.8 & \textbf{0.7} & 0.8 & 87\%\\
 Conexant & 16.6 & 17.6 & \textbf{15.9} & 106\%
 &  \textbf{1.4} & \textbf{1.4} & \textbf{1.4} & 100\%\\
 Hyperion & 18.9 & 21.5 & \textbf{18.4} & 114\%
 &  1.7 & 1.7 & \textbf{1.6} & 100\%\\
 Logitech & 3.2 & 3.7 & \textbf{3.1} & 116\%
 &  0.7 & \textbf{0.6} & 0.7 & 86\%\\
 NVidia & \textbf{6.9} & 8.6 & \textbf{6.9} & 125\%
 &  \textbf{0.9} & \textbf{0.9} & \textbf{0.9} & 100\%\\
 Palm & 24.5 & 25.9 & \textbf{22.5} & 106\%
 &  \textbf{1.8} & \textbf{1.8} & \textbf{1.8} & 100\%\\
 RedHat & \textbf{58.1} & 65.2 & 58.3 & 112\%
 &  3.8 & 3.8 & \textbf{3.6} & 100\%\\
 RSA & \textbf{7.1} & 8.7 & 7.3 & 123\%
 &  \textbf{1.2} & \textbf{1.2} & \textbf{1.2} & 100\%\\
 Sandisk & 6.5 & 7.6 & \textbf{6.4} & 117\%
 &  1.0 & 1.0 & \textbf{0.9} & 100\%\\ \hline 
\end{tabular}

\end{center}
\end{table*}


\section{ECGs and Segmentation}

\begin{figure}
 	\center\includegraphics[width=\standardlength,angle=0]{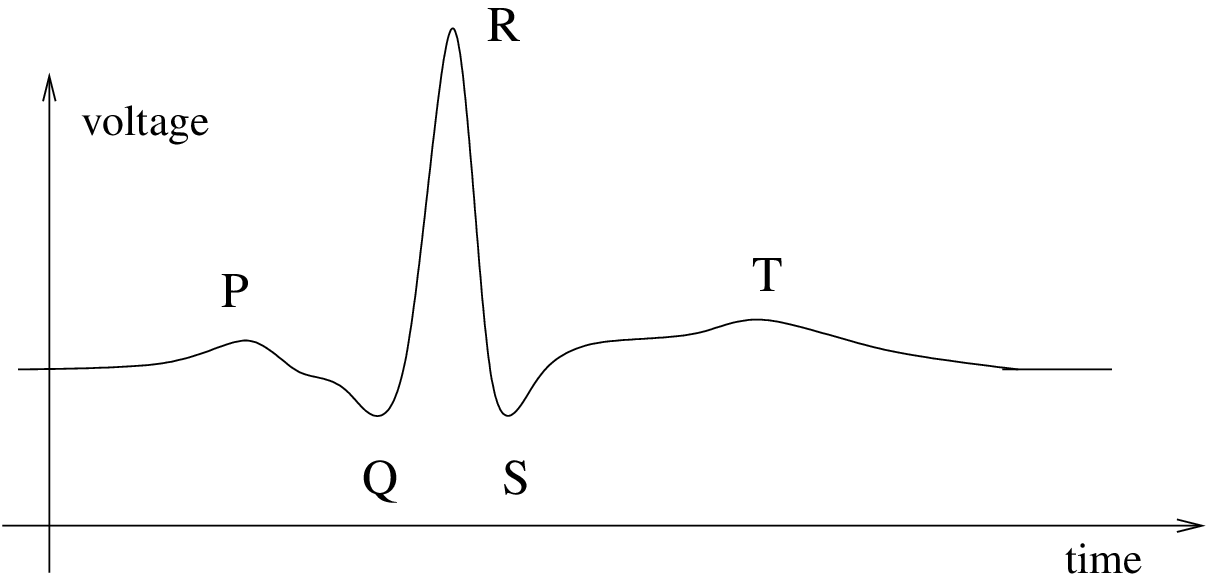}
 \caption{\label{pqrst}A typical ECG pulse with PQRST reference points. }
 \end{figure}

 Electrocardiograms (ECGs) are records of the electrical voltage in the heart. They are one
of the primary tool  in screening and diagnosis of cardiovascular diseases.
The resulting time series are nearly periodic with several commonly identifiable extrema per pulse including reference points P, Q, R, S, and T  (see Fig.~\ref{pqrst}). Each one of the extrema
has some importance:
\begin{itemize}
 \item a missing P extrema may indicate arrhythmia (abnormal heart rhythms);
 \item a large Q value may be a sign of scarring;
 \item the somewhat flat region between the S and T points is called the ST segment and its level is an indicator of ischemia~\cite{lemireieee2000}.
\end{itemize}
ECG segmentation models, including the  piecewise linear model~\cite{Tomek1974,741452}, are used for compression, monitoring or diagnosis.

 We use  ECG samples from the MIT-BIH Arrhythmia Database~\cite{PhysioNet}. The signals are recorded at a sampling rate of 360~samples per second with 11 bits resolution. Prior to segmentation, we choose time intervals spanning 300~samples (nearly 1 second) centered around the QRS complex. We select 5 such intervals by a moving window in the files of 6 different patients (``100.dat'', ``101.dat'', ``102.dat'', ``103.dat'', ``104.dat'', ``105.dat''). The model complexity varies $k=10,20,30$.

\begin{table}
\caption{\label{table-ecgerror}Comparison of top-down heuristics on ECG data  ($n=200$) for various model complexities: segmentation error
and leave-one-out cross-validation error.}
\begin{center}\small
Fit error for $k=10,20,30$.
\begin{tabular}{cccccc}
 \hline
   patient & adaptive & linear  & constant  & linear/adaptive  \\ \hline
 100 & \textbf{99.0} & 110.0 & 116.2 & 111\%\\
 101 & \textbf{142.2} & 185.4 & 148.7 & 130\%\\
  102 & \textbf{87.6} & 114.7 & 99.9 & 131\%\\
  103 & \textbf{215.5} & 300.3 & 252.0 & 139\%\\
  104 & \textbf{124.8} & 153.1 & 170.2 & 123\%\\
  105 & \textbf{178.5} & 252.1 & 195.3 & 141\%\\ \hline
  average &\textbf{141.3 }& 185.9 & 163.7 & 132\%\\
\hline
  100 & \textbf{46.8} & 53.1 & 53.3 & 113\%\\
  101 & \textbf{55.0} & 65.3 & 69.6 & 119\%\\
  102 & \textbf{42.2} & 48.0 & 50.2 & 114\%\\
  103 & \textbf{88.1} & 94.4 & 131.3 & 107\%\\
  104 & \textbf{53.4} & \textbf{53.4} & 84.1 & 100\%\\
  105 & \textbf{52.4} & 61.7 & 97.4 & 118\%\\ \hline
  average &\textbf{56.3 }& 62.6 & 81.0 & 111\%\\
\hline
  100 & \textbf{33.5} & 34.6 & 34.8 & 103\%\\
  101 & \textbf{32.5} & 33.6 & 40.8 & 103\%\\
 102 & \textbf{30.0} & 32.4 & 35.3 & 108\%\\
  103 & \textbf{59.8} & 63.7 & 66.5 & 107\%\\
  104 & \textbf{29.9} & 30.3 & 48.0 & 101\%\\
  105 & \textbf{35.6} & 37.7 & 60.2 & 106\%\\ \hline
  average &\textbf{36.9 }& 38.7 & 47.6 & 105\%\\
\hline
\end{tabular}
Leave-one-out error for $k=10,20,30$.
 
\begin{tabular}{cccccc}
 \hline
   patient & adaptive & linear  & constant  & linear/adaptive  \\ \hline
100 & \textbf{3.2} & 3.3 & 3.7 & 103\%\\
 101 & \textbf{3.8} & 4.5 & 4.3 & 118\%\\
 102 & 4.0 & 4.1 & \textbf{3.5} & 102\%\\
 103 & \textbf{4.6} & 5.7 & 5.5 & 124\%\\
 104 & 4.3 & \textbf{4.1} & 4.3 & 95\%\\
 105 & \textbf{3.6} & 4.2 & 4.5 & 117\%\\ \hline
 average &\textbf{3.9} & 4.3 & 4.3 & 110\%\\
\hline
 100 & \textbf{2.8} & \textbf{2.8} & 3.5 & 100\%\\
 101 & \textbf{3.3} & \textbf{3.3} & 3.6 & 100\%\\
 102 & 3.3 & \textbf{3.0} & 3.4 & 91\%\\
 103 & \textbf{2.9} & 3.1 & 4.7 & 107\%\\
 104 & 3.8 & 3.8 & \textbf{3.6} & 100\%\\
 105 & \textbf{2.4} & 2.5 & 3.6 & 104\%\\ \hline
 average &\textbf{3.1} & \textbf{3.1} & 3.7 & 100\%\\
\hline
 100 & 2.8 & \textbf{2.2} & 3.3 & 79\%\\
 101 & \textbf{2.9} & \textbf{2.9} & 3.6 & 100\%\\
102 & 3.3 & \textbf{2.9} & 3.3 & 88\%\\
 103 & 3.7 & \textbf{3.1} & 4.4 & 84\%\\
 104 & \textbf{3.2} & \textbf{3.2} & 3.5 & 100\%\\
 105 & \textbf{2.1} & \textbf{2.1} & 3.4 & 100\%\\ \hline
 average &3.0 & \textbf{2.7} & 3.6 & 90\%\\ \hline
\end{tabular}
\end{center}
\end{table}

\begin{figure}
  	\center\includegraphics[height=\standardlength,angle=270]{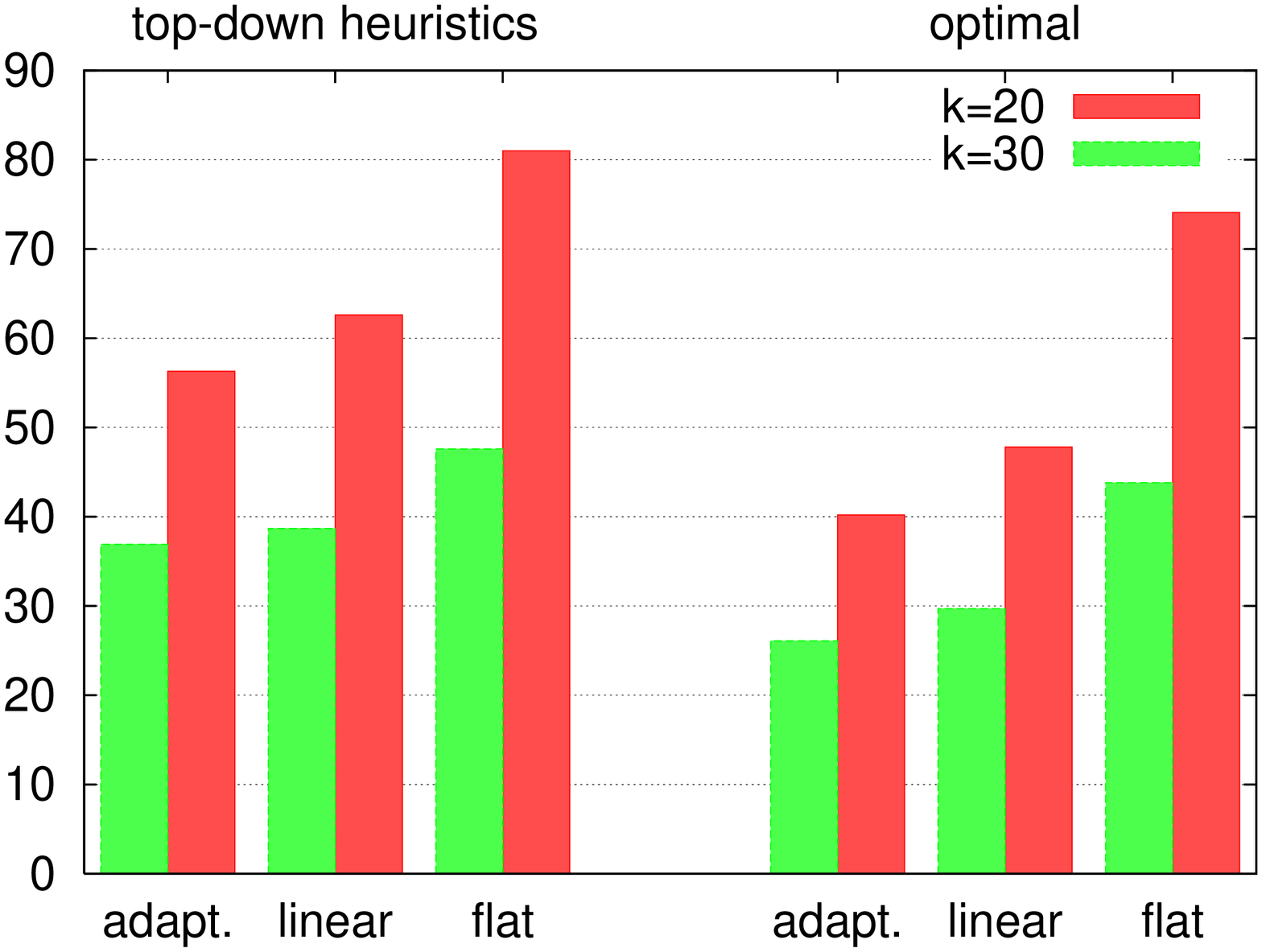}
 \caption{\label{ecgplot} Average Euclidean ($l_2$ norm) fit error over ECGs for 6~different patients. }
  \end{figure}

The segmentation error as well as the leave-one-out error are given in Table~\ref{table-ecgerror} for each patient and they are plotted in Fig.~\ref{ecgplot} in aggregated form, including the optimal errors. With the same model complexity, the adaptive top-down heuristic is better than the linear top-down heuristic~(>5\%), but more importantly,  we reduce the leave-one-out cross-validation error as well for small model complexities.
As the model complexity increases, the adaptive model eventually has a slightly worse cross-validation
error.
Unlike for the random walk and stock market data, the piecewise constant model is no longer competitive in this
case.  The optimal solution, in this case, is far more competitive with an improvement of approximately $30\%$ of the heuristics, but the relative results are the same.

%

\section{Conclusion and Future Work}

We argue that if one requires a multimodel segmentation including  flat and linear intervals,
it is better to segment accordingly instead of post-processing a piecewise linear segmentation.
Mixing drastically different interval models (monotonic and linear) 
and offering richer, more flexible segmentation models 
remains an important open problem.

To ease comparisons accross different models, we propose a simple complexity model based
on counting the number of regressors.
As  supporting evidence that mixed models are competitive, we consistently improved the accuracy 
by 5\% and 13\% respectively without increasing the cross-validation error  over white noise and random walk data.
Moreover, whether we consider stock market prices of ECG data, for small model complexity, the adaptive
top-down heuristic is noticeably better than the commonly used top-down linear heuristic.
The adaptive segmentation heuristic
 is not significantly harder to implement nor slower than the top-down linear heuristic.

We proved that optimal adaptive time series segmentations can be
computed in quadratic time, when the model complexity and the
polynomial degree are small.
However, despite this low complexity, optimal segmentation
by dynamic programming is not an option for real-world time series (see Fig.~\ref{timings}).
With reason, some researchers go as far as
not even discussing dynamic programming as an alternative~\cite{657889}.
In turn, we have shown that adaptive top-down heuristics can be implemented in linear time
 after the linear time computation of a buffer. In our experiments, for a small model complexity,
the top-down heuristics are competitive with the dynamic programming alternative which sometimes
offer small gains ($10\%$).

Future work will investigate real-time processing for online applications such as high frequency trading~\cite{1007574} and live patient
monitoring. An ``amnesic'' approach should be tested~\cite{palpanas04online}.


\section{Acknowledgments}

The author 
wishes to thank Owen Kaser of UNB, Martin Brooks of NRC, 
 and Will Fitzgerald of NASA Ames for their insightful
comments.

\bibliographystyle{plain} 
\bibliography{episode,multidegree} 

\begin{thebibliography}{10}

\bibitem{yahoofinance}
{Yahoo! Finance}.
\newblock last accessed June, 2006.

\bibitem{lowellreport}
S.~Abiteboul, R.~Agrawal, et~al.
\newblock {The Lowell Database Research Self Assessment}.
\newblock Technical report, Microsoft, 2003.

\bibitem{allen1974rbv}
D.~M. Allen.
\newblock {The Relationship between Variable Selection and Data Agumentation
  and a Method for Prediction}.
\newblock {\em Technometrics}, 16(1):125--127, 1974.

\bibitem{507653}
S.~Anand, W.-N. Chin, and S.-C. Khoo.
\newblock Charting patterns on price history.
\newblock In {\em ICFP'01}, pages 134--145, New York, NY, USA, 2001. ACM Press.

\bibitem{atkeson1997lwl}
C.~G.~F. Atkeson, A.~W.~F. Moore, and S.~F. Schaal.
\newblock {Locally Weighted Learning}.
\newblock {\em Artificial Intelligence Review}, 11(1):11--73, 1997.

\bibitem{Balvers2000}
R.~Balvers, Y.~Wu, and E.~Gilliland.
\newblock Mean reversion across national stock markets and parametric
  contrarian investment strategies.
\newblock {\em Journal of Finance}, 55:745--772, 2000.

\bibitem{bellman1969}
R.~Bellman and R.~Roth.
\newblock Curve fitting by segmented straight lines.
\newblock {\em J. Am. Stat. Assoc.}, 64:1079--1084, 1969.

\bibitem{birattari1999llm}
M.~Birattari, G.~Bontempi, and H.~Bersini.
\newblock {Lazy learning for modeling and control design}.
\newblock {\em Int. J. of Control}, 72:643--658, 1999.

\bibitem{363580}
H.~J. Breaux.
\newblock {A modification of {Efroymson}'s technique for stepwise regression
  analysis}.
\newblock {\em Commun. ACM}, 11(8):556--558, 1968.

\bibitem{YLBIJCAI05}
M.~Brooks, Y.~Yan, and D.~Lemire.
\newblock Scale-based monotonicity analysis in qualitative modelling with flat
  segments.
\newblock In {\em IJCAI'05}, 2005.

\bibitem{Burnham2004}
K.~P. Burnham and D.~R. Anderson.
\newblock Multimodel inference: understanding aic and bic in model selection.
\newblock In {\em {Amsterdam Workshop on Model Selection}}, 2004.

\bibitem{343930}
S.~Chardon, B.~Vozel, and K.~Chehdi.
\newblock Parametric blur estimation using the generalized cross-validation.
\newblock {\em Multidimensional Syst. Signal Process.}, 10(4):395--414, 1999.

\bibitem{Chaudhuri2003}
K.~Chaudhuri and Y.~Wu.
\newblock Random walk versus breaking trend in stock prices: Evidence from
  emerging markets.
\newblock {\em Journal of Banking \& Finance}, 27:575--592, 2003.

\bibitem{Donoho1994}
D.~L. Donoho and I.~M. Johnstone.
\newblock Ideal spatial adaptation by wavelet shrinkage.
\newblock {\em Biometrika}, 81:425--455, 1994.

\bibitem{Fama1988}
E.~F. Fama and K.~R. French.
\newblock Permanent and temporary components of stock prices.
\newblock {\em Journal of Political Economy}, 96:246--273, 1988.

\bibitem{AAAI05}
W.~Fitzgerald, D.~Lemire, and M.~Brooks.
\newblock Quasi-monotonic segmentation of state variable behavior for reactive
  control.
\newblock In {\em AAAI'05}, 2005.

\bibitem{Foster1994}
D.~P. Foster and E.~I. George.
\newblock The risk inflation criterion for multiple regression.
\newblock {\em Annals of Statistics}, 22:1947--1975, 1994.

\bibitem{Friedl2002}
H.~Friedl and E.~Stampfer.
\newblock {\em Encyclopedia of Environmetrics}, chapter Cross-Validation.
\newblock Wiley, 2002.

\bibitem{Friedman1991}
J.~Friedman.
\newblock Multivariate adaptive regression splines.
\newblock {\em Annals of Statistics}, 19:1--141, 1991.

\bibitem{fu2004fts}
T.~Fu, F.~Chung, R.~Luk, and C.~Ng.
\newblock {Financial Time Series Indexing Based on Low Resolution Clustering}.
\newblock {\em ICDM-2004 Workshop on Temporal Data Mining}, 2004.

\bibitem{citeulike:799461}
D.~G. Galati and M.~A. Simaan.
\newblock Automatic decomposition of time series into step, ramp, and impulse
  primitives.
\newblock {\em Pattern Recognition}, 39(11):2166--2174, November 2006.

\bibitem{PhysioNet}
A.~L. Goldberger, L.~A.~N. Amaral, et~al.
\newblock {PhysioBank, PhysioToolkit, and PhysioNet}.
\newblock {\em Circulation}, 101(23):215--220, 2000.

\bibitem{Haiminen04}
N.~Haiminen and A.~Gionis.
\newblock Unimodal segmentation of sequences.
\newblock In {\em ICDM'04}, 2004.

\bibitem{han98}
J.~Han, W.~Gong, and Y.~Yin.
\newblock Mining segment-wise periodic patterns in time-related databases.
\newblock In {\em KDD'98}, 1998.

\bibitem{Hastie2001}
T.~Hastie, R.~Tibshirani, and J.~Friedman.
\newblock {\em Elements of Statistical Learning: Data Mining, Inference and
  Prediction}.
\newblock Springer-Verlag, 2001.

\bibitem{kamruzzaman2003sbm}
J.~Kamruzzaman, RA~Sarker, and I.~Ahmad.
\newblock {SVM based models for predicting foreign currency exchange rates}.
\newblock {\em ICDM 2003}, pages 557--560, 2003.

\bibitem{657889}
E.~J. Keogh, S.~Chu, D.~Hart, and M.~J. Pazzani.
\newblock An online algorithm for segmenting time series.
\newblock In {\em ICDM'01}, pages 289--296, Washington, DC, USA, 2001. IEEE
  Computer Society.

\bibitem{keogh2003nts}
E.~J. Keogh and S.~F. Kasetty.
\newblock {On the Need for Time Series Data Mining Benchmarks: A Survey and
  Empirical Demonstration}.
\newblock {\em Data Mining and Knowledge Discovery}, 7(4):349--371, 2003.

\bibitem{KeoghP98}
E.~J. Keogh and M.~J. Pazzani.
\newblock An enhanced representation of time series which allows fast and
  accurate classification, clustering and relevance feedback.
\newblock In {\em KDD'98}, pages 239--243, 1998.

\bibitem{kleinberg2006ad}
J.~Kleinberg and {\'E}.~Tardos.
\newblock {\em {Algorithm design}}.
\newblock Pearson/Addison-Wesley, 2006.

\bibitem{LemireCASCON2002}
D.~Lemire.
\newblock Wavelet-based relative prefix sum methods for range sum queries in
  data cubes.
\newblock In {\em CASCON 2002}. IBM, October 2002.

\bibitem{YLBICDMO05}
D.~Lemire, M.~Brooks, and Y.~Yan.
\newblock An optimal linear time algorithm for quasi-monotonic segmentation.
\newblock In {\em ICDM'05}, 2005.

\bibitem{Ola2004}
D.~Lemire and O.~Kaser.
\newblock Hierarchical bin buffering: Online local moments for dynamic external
  memory arrays.
\newblock submitted in February 2004.

\bibitem{lemireieee2000}
D.~Lemire, C.~Pharand, et~al.
\newblock Wavelet time entropy, t wave morphology and myocardial ischemia.
\newblock {\em IEEE Transactions in Biomedical Engineering}, 47(7), July 2000.

\bibitem{Lindstrom99}
M.~Lindstrom.
\newblock Penalized estimation of free-knot splines.
\newblock {\em Journal of Computational and Graphical Statistics}, 1999.

\bibitem{lo1988smp}
A.~W. Lo and A.~C. MacKinlay.
\newblock {Stock Market Prices Do Not Follow Random Walks: Evidence from a
  Simple Specification Test}.
\newblock {\em The Review of Financial Studies}, 1(1):41--66, 1988.

\bibitem{Zuo97}
Z.~Luo and G.~Wahba.
\newblock Hybrid adaptive splines.
\newblock {\em Journal of the American Statistical Association}, 1997.

\bibitem{GaetanMonari06012002}
G.~Monari and G.~Dreyfus.
\newblock {Local Overfitting Control via Leverages}.
\newblock {\em Neural Comp.}, 14(6):1481--1506, 2002.

\bibitem{palpanas04online}
T.~Palpanas, T.~Vlachos, E.~Keogh, D.~Gunopulos, and W.~Truppel.
\newblock Online amnesic approximation of streaming time series.
\newblock In {\em ICDE 2004}, 2004.

\bibitem{Pednault1991}
E.~Pednault.
\newblock Minimum length encoding and inductive inference.
\newblock In G.~Piatetsky-Shapiro and W.~Frawley, editors, {\em Knowledge
  Discovery in Databases}. AAAI Press, 1991.

\bibitem{sornette2002smc}
D.~Sornette.
\newblock {\em {Why Stock Markets Crash: Critical Events in Complex Financial
  Systems}}.
\newblock Princeton University Press, 2002.

\bibitem{DBLP:conf/sdm/TerziT06}
E.~Terzi and P.~Tsaparas.
\newblock Efficient algorithms for sequence segmentation.
\newblock In {\em SDM'06}, 2006.

\bibitem{Tomek1974}
I.~Tomek.
\newblock Two algorithms for piecewise linear continuous approximations of
  functions of one variable.
\newblock {\em IEEE Trans. on Computers}, C-23:445--448, April 1974.

\bibitem{tse2002mai}
P.~Tse and J.~Liu.
\newblock {Mining Associated Implication Networks: Computational Intermarket
  Analysis}.
\newblock {\em ICDM 2002}, pages 689--692, 2002.

\bibitem{Tsuda2000}
K.i Tsuda, G.~R\"atsch, S.~Mika, and K.-R. M\"uller.
\newblock Learning to predict the leave-one-out error.
\newblock In {\em {NIPS*2000} Workshop: Cross-Validation, Bootstrap and Model
  Selection}, 2000.

\bibitem{vasko2002}
K.~T. Vasko and H.~T.~T. Toivonen.
\newblock Estimating the number of segments in time series data using
  permutation tests.
\newblock In {\em ICDM'02}, 2002.

\bibitem{741452}
H.~J. L.~M. Vullings, M.~H.~G. Verhaegen, and H.~B. Verbruggen.
\newblock {ECG} segmentation using time-warping.
\newblock In {\em IDA'97}, pages 275--285, London, UK, 1997. Springer-Verlag.

\bibitem{1007574}
H.~Wu, B.~Salzberg, and D.~Zhang.
\newblock Online event-driven subsequence matching over financial data streams.
\newblock In {\em SIGMOD'04}, pages 23--34, New York, NY, USA, 2004. ACM Press.

\bibitem{zhu2003qht}
Y.~Zhu and D.~Shasha.
\newblock {Query by humming: a time series database approach}.
\newblock {\em Proc. of SIGMOD}, 2003.

\end{thebibliography}


%
\balance

\end{document}